\documentclass[useAMS,usenatbib]{mn2e}
\usepackage{amsmath}
\usepackage{graphicx}
\usepackage{epstopdf}
\usepackage{amssymb,latexsym}
\usepackage{multicol}
\usepackage{color}

\begin{document}

\title[Discovery of new open cluster Cepheids in the VVV survey]{New
  open cluster Cepheids in the VVV survey tightly constrain
  near-infrared period--luminosity relations}

\author[Xiaodian Chen, Richard de Grijs and Licai Deng] {Xiaodian
  Chen$^{1,2}$, Richard de Grijs$^{1,3}$ and Licai
  Deng$^{2}$\\
$^{1}$Kavli Institute for Astronomy \& Astrophysics and
  Department of Astronomy, Peking University, Yi He Yuan Lu 5, Hai
  Dian\\ District, Beijing 100871, China\\
$^{2}$Key Laboratory for Optical Astronomy, National Astronomical
  Observatories, Chinese Academy of Sciences, 20A Datun Road,\\ Chaoyang
  District, Beijing 100012, China\\
$^{3}$International Space Science Institute--Beijing, 1 Nanertiao,
  Zhongguancun, Hai Dian District, Beijing 100190, China}

\date{xxx}

\pagerange{\pageref{firstpage}--\pageref{lastpage}} \pubyear{2016}
\label{firstpage}

\maketitle

\begin{abstract}
Classical Cepheids are among the most useful Galactic and nearby
extragalactic distance tracers because of their well-defined
period--luminosity relations (PLRs). Open cluster (OC) Cepheids are
important objects to independently calibrate these PLRs. Based on Data
Release 1 of the {\sl VISTA} Variables in the V\'ia L\'actea survey,
we have discovered four new, faint and heavily reddened OC Cepheids,
including the longest-period OC Cepheid known, ASAS J180342$-$2211.0
in Teutsch 14a. The other OC--Cepheid pairs include NGC 6334 and V0470
Sco, Majaess 170 and ASAS J160125$-$5150.3, and Teutsch 77 and BB
Cen. ASAS J180342$-$2211.0, with a period of $\log P = 1.623$ [days]
is important to constrain the slope of the PLR. The currently most
complete $JHK_{\rm s}$ Galactic Cepheid PLRs are obtained based on a
significantly increased sample of 31 OC Cepheids, with associated
uncertainties that are improved by 40 per cent compared with previous
determinations (in the $J$ band). The NIR PLRs are in good
  agreement with previous PLRs determined based on other methods.
\end{abstract}

\begin{keywords}
methods: data analysis --- stars: variables: Cepheids ---
  open clusters and associations: general --- stars: distances
\end{keywords}

\section{Introduction}

Classical Cepheids are among the most useful Galactic and nearby
extragalactic distance tracers because of their well-established
period--luminosity relations \citep[PLRs; also known as the `Leavitt
  law':][]{Leavitt12}. Cepheids are commonly used to anchor
extragalactic distances, constrain the Hubble constant and study
Galactic structure. They also relate directly to secondary distance
tracers, such as Type Ia supernovae, whose luminosities are calibrated
using Cepheids in their host galaxies \citep{Riess11}. The Carnegie
Hubble Program \citep{Freedman11} aims at determining an accurate
Hubble constant based on the Cepheid distance scale,
\citet{Matsunaga11, Matsunaga15}, \citet{Feast14} and \citet{Dekany15}
have studied the Milky Way's structure in the Galactic bulge based on
Cepheids, while \citet{Majaess09} and \citet{Dambis15} mapped the
Galactic spiral arms using distances to classical Cepheids.

Cepheid PLRs are not only important for measuring distances but also
to constrain the pulsation physics and the evolution of
Cepheids. However, independent Cepheid distances must be obtained,
which can be achieved using (i) trigonometric parallaxes
\citep{Feast97, Benedict07}, (ii) Baade--Wesselink-type
methods/surface-brightness techniques \citep{Gieren97, Storm11} and
(iii) main-sequence or isochrone fitting based on open cluster (OC)
photometry \citep{An07, Turner10, Anderson13, Chen15}. The OC
parameters derived based on the latter method can be used as
independent constraints on the distances, ages and reddening values of
Cepheids in these clusters. This method can also be used to study
Cepheids in heavily reddened environments.

The key to establishing the PLR using OC Cepheids is to select
high-confidence cluster Cepheids. \citet{An07} obtained $BVI_{\rm
  c}JHK_{\rm s}$ PLRs based on seven Galactic OCs hosting Cepheids,
while \citet{Anderson13} found five OC Cepheids using an
eight-dimensional selection approach and obtained a $V$-band PLR for
18 OC Cepheids. \citet{Chen15} found six new OC Cepheids based on
near-infrared (NIR) data and derived a $J$-band PLR for 19 OC
Cepheids. As more accurate data are becoming available, the number of
OC Cepheids is increasing, whereas the confidence of OC membership is
improving simultaneously. \citet{Majaess11} provided new evidence to
support the likely membership of the Cepheid TW Nor of the OC
Lyng{\aa} 6. \citet{Turner12} discovered that SU Cas is a probable
member of the OC Alessi 95; \citet{Majaess12a} conducted a detailed
distance analysis of this latter OC and its member Cepheids based on
new X-ray and $JHK_{\rm s}$ data. \citet{Majaess13} assessed the links
between three Cepheids and NGC 6067 based on data from the {\sl VISTA}
Variables in the V\'ia L\'actea (VVV) survey. However, the number of
OC Cepheids that can presently be used to establish OC--Cepheid PLRs
is still small, of order 20. One limitation to further progress is
that approximately half of the OCs in the DAML02 catalogue
\citep{Dias02} are poorly studied. Another limitation is a lack of
availability of mean Cepheid intensities in multiple passbands;
indeed, for many Cepheids we only have access to their $V$-band light
curves.

In this paper, we use VVV Data Release 1 (DR1)
\citep{Minniti10,Saito12} to study faint OCs in the Galactic bulge and
the Galactic midplane. We improve the quality of their age, reddening
and distance determinations. We also convert single-epoch NIR
photometry to mean intensities based on application of light-curve
templates. We have collected the current largest sample of 31
high-confidence OC Cepheids based on NIR photometry, thus enabling us
to obtain the most accurate $JHK_{\rm s}$ PLRs for Galactic OC
Cepheids to date. In fact, in this paper we derive the first
statistically meaningful OC--Cepheid $H$- and $K_{\rm s}$-band PLRs;
our updated $J$-band PLR represents a significant improvement with
respect to our earlier derivation \citep{Chen15}. In section 2 we
present the data, the method used and our OC--Cepheid
catalogue. Section 3 discusses the OC membership characteristics of
our four newly found Cepheids. In Section 4 we discuss the Cepheid
$JHK_{\rm s}$ PLRs, while Section 5 summarizes and concludes this
paper.

\section{Method}

NIR $JHK_{\rm s}$ photometry is suitable for studies of young OCs, not
only because of the small or even negligible effects expected owing to
differential extinction, but also in the context of detecting heavily
obscured main-sequence stars \citep{Majaess12b}. Using the Two Micron
All-Sky Survey (2MASS) Point Source Catalog \citep{Cutri03}, thousands
of OC parameters can be determined easily using main-sequence
fitting. Complete samples can reach distances of order 1.8 kpc
\citep{Kharchenko13}. With the availability of NIR surveys fainter
than 2MASS, such as the UKIRT Infrared Deep Sky Survey (UKIDSS) DR6
Galactic Plane Survey \citep{Lucas08} and VVV DR1, even more distant,
heavily reddened OC properties can be determined. The VVV survey is a
$ZYJHK_{\rm s}$ photometric survey focussed on the Galactic Centre and
a few fields along the Galactic plane. It reaches four magnitudes
fainter than 2MASS. Combining 2MASS $JHK_{\rm s}$ data and VVV data,
the available NIR catalogues cover a range of more than 10 magnitudes,
thus allowing us to study Galactic stellar populations from bright
Cepheids to faint main-sequence dwarf stars.

Cepheids are young stars with ages near $\log (t \mbox{ yr}^{-1}) =
8.0$, so they can be present in young OCs. To find OC Cepheids, we
rely on properties such as their spatial positions, distances and
proper motions, as well as their radial velocities, ages and
metallicities \citep{Anderson13,Chen15}. Usually, the positions and
distances are easily obtained, while proper-motion and radial-velocity
data are only available for smaller numbers of specific OCs. To define
our OC master sample, we included not only the 2167 OCs in the January
2016 version of the DAML02 catalogue, but also some other new OCs
recently found by \citet{Borissova11, Borissova14}, \citet{Chene12},
\citet{Schmeja14} and \citet{Scholz15}. In addition, we collected all
Cepheids in the All-Sky Automated Survey (ASAS) Catalogue of Variable
Stars \citep{Pojmanski02}, the General Catalogue of Variable Stars
(GCVS) and \citet{Tammann03}. Positional cross-matching within 1
degree (projected) separation in the VVV field yielded 203 potential
OC--Cepheid pairs. In our next selection step we excluded Cepheids
that are located more than three times their host OC's apparent radius
from the cluster centre, leaving 22 OCs for further analysis.

\begin{figure}
\centering
\includegraphics[width=80mm]{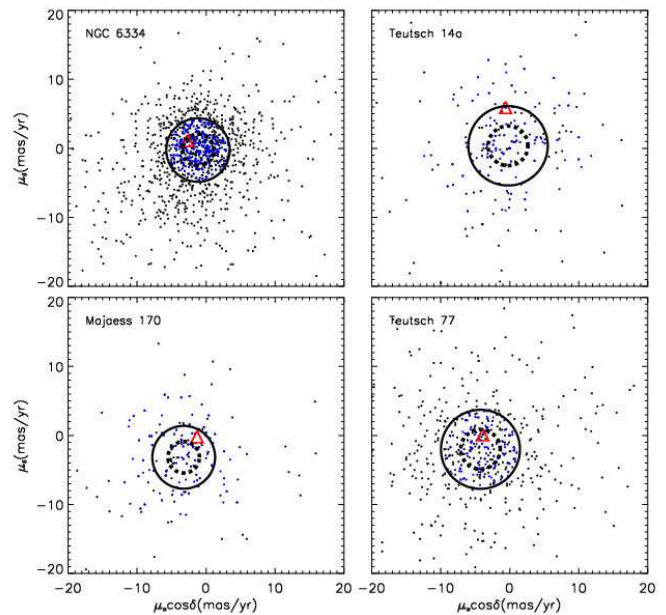}
\caption{Proper-motion selection of our four target OCs. Solid and
  dotted circles show, respectively, the 1$\sigma$ and 0.5$\sigma$
  radii of the clusters' proper motions. Black dots: stars with
  uncertainties smaller than our error cut of 15 mas yr$^{-1}$ (the
  typical 1$\sigma$ uncertainty in the proper-motion catalogue is 4--5
  mas yr$^{-1}$). Blue dots: cluster members located inside the
  1$\sigma$ (well-populated OCs) or 2$\sigma$ radii (sparsely
  populated OCs). Red triangles: Cepheid proper motions.}
  \label{c2f1.fig}
\end{figure}

To obtain stellar samples of genuine OC members, we next proceeded
with proper-motion analysis. Proper-motion data were obtained from the
the Fourth US Naval Observatory CCD Astrograph Catalog
\citep[UCAC4;][]{Zacharias13}. Stars with proper-motion errors in
excess of 20 mas yr$^{-1}$ were excluded from our initial
samples. Subsequently, we calculated the average and standard
deviation for all stars in the distribution so as to exclude
high-proper-motion stars, adopting a 1$\sigma$ selection cut. This
process was repeated twice to obtain reliable proper motions for each
cluster. All stars located within the 1$\sigma$ proper-motion
distribution were adopted as cluster members, while for some OCs with
fewer than 100 stars within the 1$\sigma$ distribution, we selected
all stars inside the 2$\sigma$ distribution instead. We then selected
Cepheids located within the 1$\sigma$ proper-motion distribution of
their host clusters for further analysis: see Fig. \ref{c2f1.fig}. To
select genuine cluster members, we plotted the ($J, J-H$)
colour--magnitude diagrams (CMDs), combining 2MASS and VVV data. VVV
$JHK_{\rm s}$ magnitudes were converted to the 2MASS $JHK_{\rm s}$
system following the prescriptions recommended by the Cambridge
Astronomy Survey
Unit\footnote{http://casu.ast.cam.ac.uk/surveys-projects/vista/technical/\\ photometricproperties}.

For clusters exhibiting obvious main sequences, distances, reddening
values and ages can be determined by fitting the cluster sequences
with Padova isochrones \citep{Girardi00, Bressan12}. Since OCs
generally belong to a relatively young Milky Way population, their
metallicities are comparably high, on the order of the solar value. In
addition, metallicity-dependent variations in the derived cluster
parameters, including the ages and distances, are small
\citep[e.g.,][]{Kharchenko13}. Hence, we adopted solar
metallicity. The scatter in the cluster sequences is mainly caused by
differential reddening and the consequently faint stellar
luminosities. We applied our isochrone fitting in an automated fashion
to determine the smallest residuals pertaining to the parameter space
defined by the cluster ages, reddening values and distance
moduli.

  For a given isochrone defined by a specific age, reddening and
  distance modulus, our fitting procedure consisted of four
  steps. First, we linearly resampled (interpolated) the theoretical
  isochrone such that the intervals between two adjacent points were
  about 0.01 mag. Second, we selected the nearest isochrone point to
  every observed data point and determined the distance between each
  pair of points thus selected. Based on the mean distance and
  standard deviation, obvious field stars beyond the distribution's
  1$\sigma$ range were excluded. Third, we calculated the mean
  distance and standard deviation for this initial, cleaned data
  set. For bright stars, $J<13.5$ mag, we adopted stars located within
  the 2$\sigma$ range as cluster members; for fainter stars, the
  selection criterion was established independently for every 0.1 mag
  bin between $J=13.5$ mag and $J=19.0$ mag. The standard deviation
  increases from bright to faint magnitudes. Stars found within the
  1$\sigma$ range in each bin were selected as high-probability OC
  members. These cluster members are shown in the rightmost panel of
  Fig. \ref{c2f2.fig}. Finally, we calculated the uncertainty in the
  colour excess, the error in the apparent distance modulus and the
  overall uncertainty.

  Next, we allowed the age (0.1 dex age resolution), reddening (0.01
  mag resolution) and distance modulus (0.1 mag resolution) to vary
  freely in order to find the best fit, i.e., the result defined by
  the smallest overall error. Figure \ref{c2f2.fig} shows, from left
  to right and for all four newly found OCs hosting Cepheids, the
  original CMDs, the control-field CMDs, the cleaned CMDs and the CMDs
  resulting from our automated fitting. The original CMDs are composed
  of stars located in the cluster regions and based on 2MASS and VVV
  data. The control fields are located at least twice the cluster
  radius from the cluster centre, and the size of the control field is
  the same as that of the cluster field. Since the VVV fields are very
  well populated, the control fields could be used to statistically
  correct for field stars. The cleaned CMDs show the candidate cluster
  members, resulting from having statistically subtracted the control
  fields from the cluster fields. They exhibit obvious cluster
  sequences from the saturation magnitude ($J \sim 12.5$ mag) to $J
  \sim 17.0$ mag. These sequences are comparable to the sequences
  resulting from our automated fits (see the rightmost panels in
  Fig. \ref{c2f2.fig}). This confirms that the cluster sequences
  selected by our automated fits are statistically reliable. We
    did not fit the field-star-subtracted sequence directly, because
    the reliable range of object magnitudes is small, and so is the
    number of sources in the subtracted sequence. Since the VVV survey
    images were taken with very short exposure times, it is more
    appropriate (and more statistically meaningful) to perform
    statistical analyses of large samples instead of analyses based on
    a smaller number of specific, extracted sources. The basic
  physical OC parameters thus obtained are included in Table
  \ref{c2table2}. They form the basis for our OC--Cepheid
  selection. Upon application of all selection criteria, four
  high-confidence OC--Cepheid pairs remain: see Fig. \ref{c2f2.fig},
  where we show the approximate locations of the clusters' instability
  strips. OC Cepheids must be located inside the instability strip of
  their host cluster.

\begin{figure*}
\centering
  \begin{minipage}{160mm}
  \includegraphics[width=160mm]{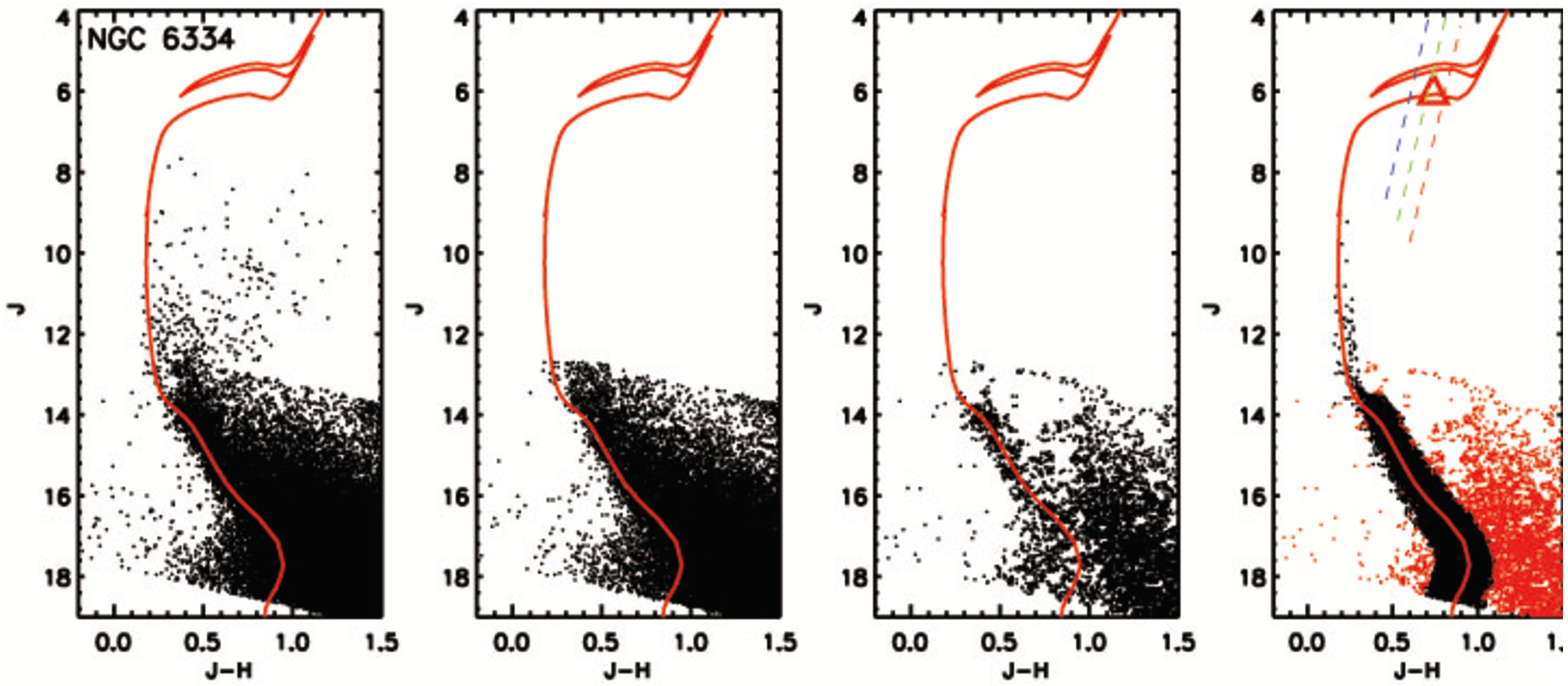}
  \includegraphics[width=160mm]{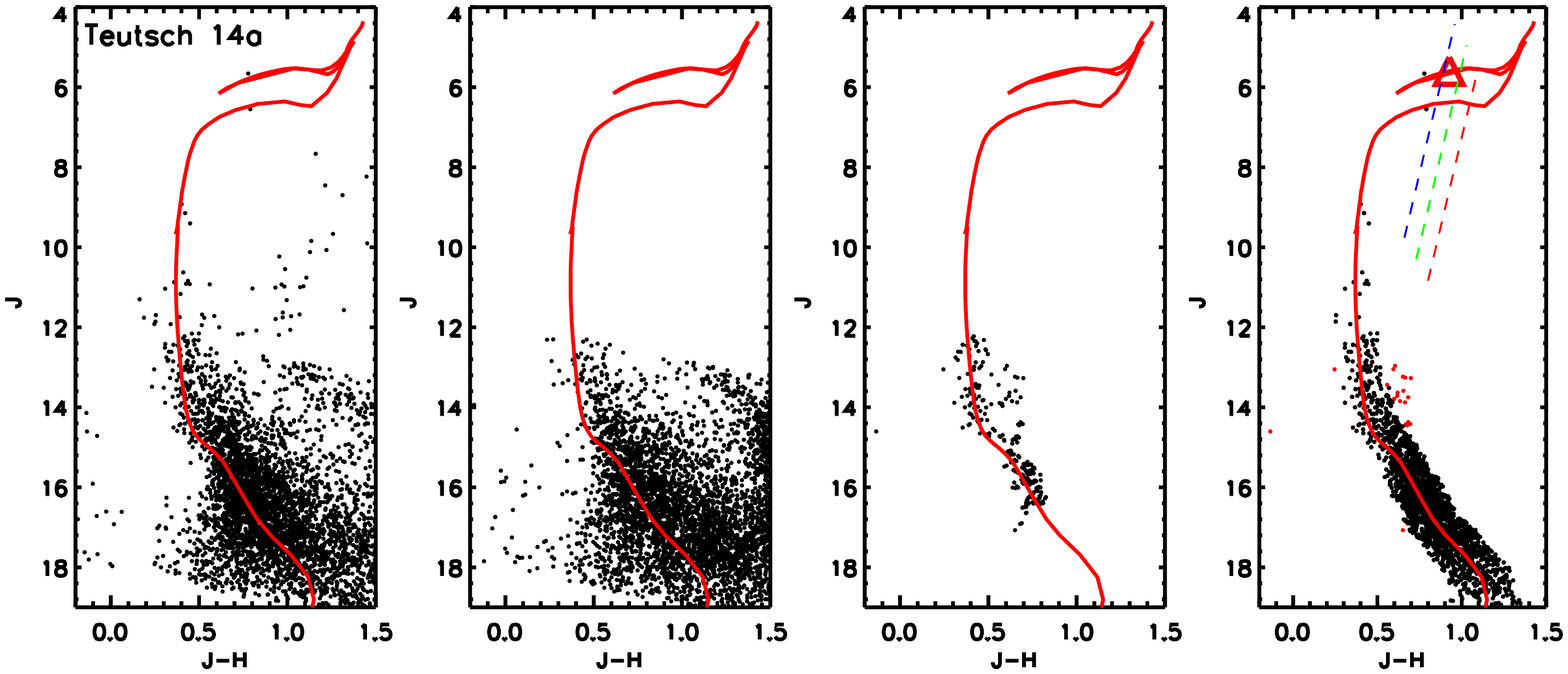}
\caption{Isochrone fits to the four OC CMDs hosting Cepheids.
    From left to right: Original CMDs, control-field CMDs,
    statistically cleaned CMDs and the CMDs resulting from our
    automated fits. In the original CMDs, the black data points are
    stars from the combined 2MASS and VVV data sets; in the
    control-field CMDs, the black data points are stars in the VVV
    control field. The cleaned CMDs show candidate cluster members,
    selected by statistically subtracting the control-field CMD from
    the cluster CMD. The black data points in the CMDs resulting from
    our automated fits are high-probability OC members; red dots:
    stars in the cleaned CMDs used to compare with the black dots.
    Solid red lines: best-fitting isochrones for solar metallicity and
    different distance moduli, reddening values and ages based on the
    Padova isochrones \citep{Girardi00, Bressan12}. Green dashed
    lines: central ridgelines of the Cepheid instability strips
    converted from \citet{Sandage08}; red and blue dashed lines: red
    and blue edges of the instability strips, respectively. The
    Cepheids' $JH$ magnitudes have been converted to mean-intensity
    magnitudes (see Section 3.3); they are shown as red
    triangles.\label{c2f2.fig}}
  \end{minipage}
\end{figure*}

\addtocounter{figure}{-1}
\begin{figure*}
  \includegraphics[width=160mm]{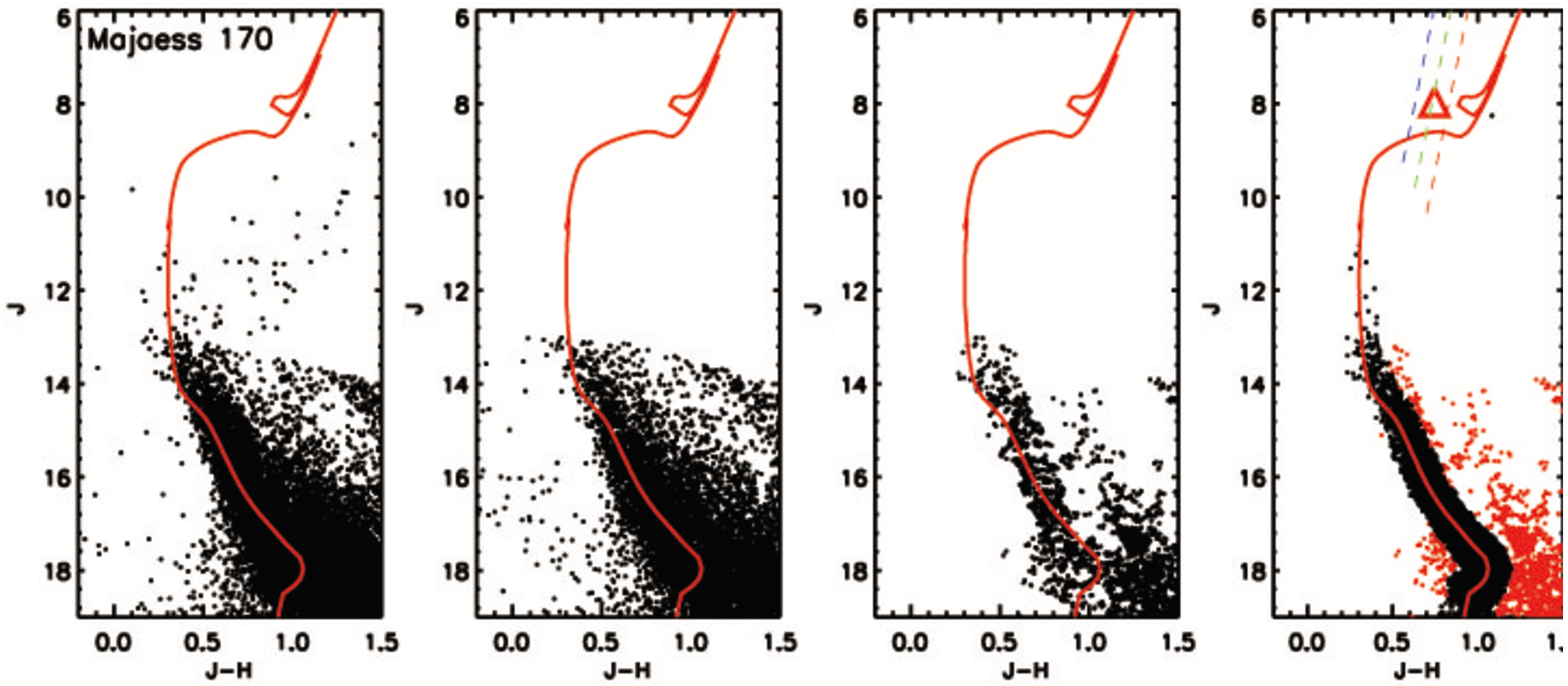}
  \includegraphics[width=160mm]{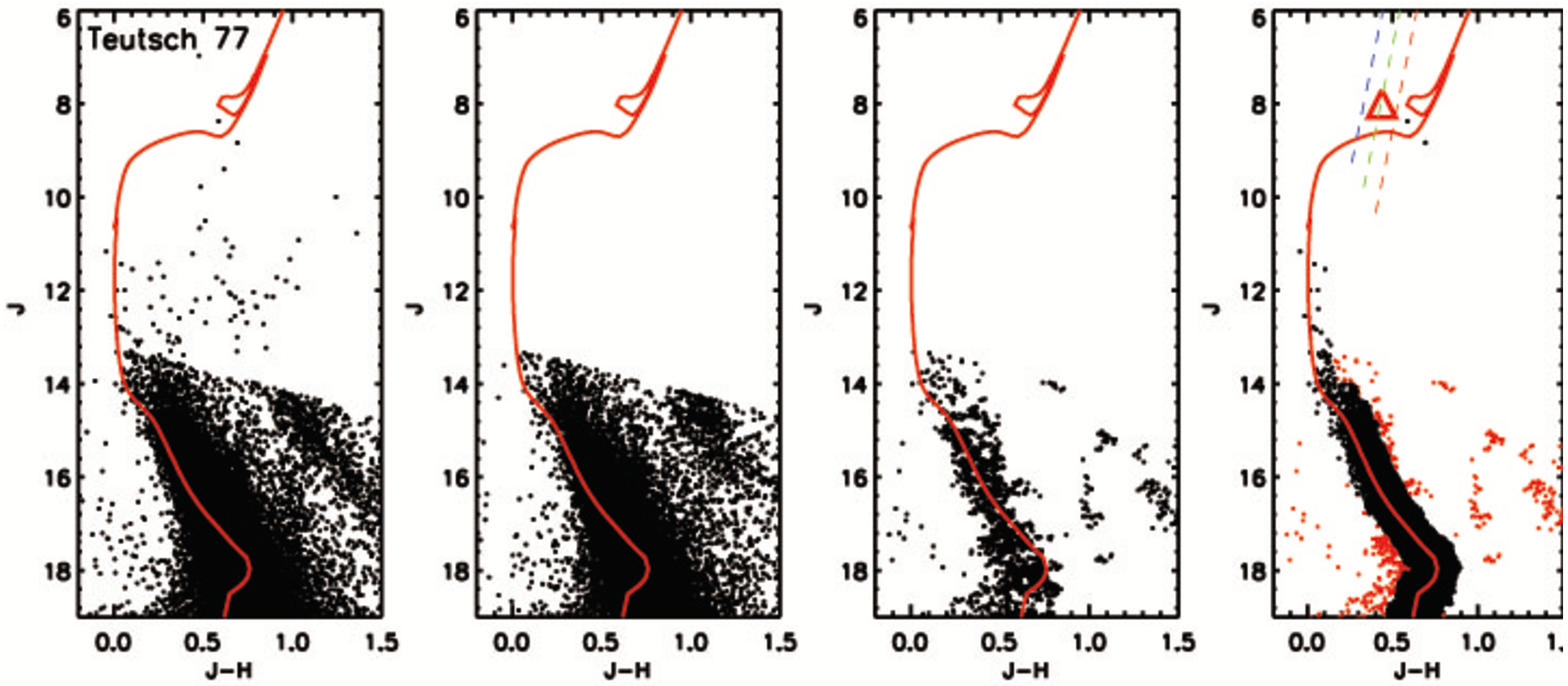}
\caption{(Continued)}
\end{figure*}

\begin{table*}
 \begin{minipage}{200mm}
\caption{Parameters of the four newly discovered VVV OCs. RV: radial velocity.\label{c2table2}}
\begin{tabular}{@{}lccccccccc@{}}
  \hline
ID & RA(J2000) & DEC(J2000) & Radius & Distance & $E(J-H)$  & $\log( t )$ & $\mu_{\alpha}$ &$\mu_{\delta}$& RV \\
   & (hh:mm:ss) & ($^{\circ}$ $'$ $''$) & (arcmin) & (kpc) & (mag) & [yr] & (mas yr$^{-1}$) & (mas yr$^{-1}$) & (km s$^{-1}$) \\
  \hline

Teutsch 14a & 18:03:31.4 & $-$22:07:31.3&5.0  & $2.0 \pm 0.2$ & $0.48 \pm 0.01$ & $7.5\pm0.3$ & $-0.2 \pm 3.9$ & $ 0.4 \pm 4.2$ &\\
Majaess 170 & 16:00:52.0 & $-$51:42:36.6&6.0  & $1.8 \pm 0.2$ & $0.38 \pm 0.03$ & $8.1\pm0.2$ & $-3.2 \pm 3.1$ & $-3.1 \pm 3.3$ &\\
Teutsch 77  & 11:53:15.6 & $-$62:36:33.2&10.0 & $2.6 \pm 0.3$ & $0.10 \pm 0.02$ & $8.1\pm0.2$ & $-4.2 \pm 4.2$ & $-2.0 \pm 3.8$ &\\
NGC 6334    & 17:20:52.7 & $-$36:07:02.5&20.0 & $1.7 \pm 0.2$ & $0.28 \pm 0.02$ & $7.7\pm0.3$ & $-1.2 \pm 3.1$ & $-0.2 \pm 3.4$ &$-4\pm2$\\

\hline
 \end{tabular}
\end{minipage}
\end{table*}

\section{Results: individual open cluster--Cepheid matches}

The four newly found OC Cepheids are more distant long-period
Cepheids, hidden behind heavier extinction, than the OC Cepheids found
previously. Their inclusion has the potential to improve the
importance of the NIR PLRs' calibrations based on the OC-matching
method. In addition, long-period Cepheids (few of which are available
to date) are important to accurately determine the slope of the PLRs.

NGC 6334 is a heavily reddened star-forming region. The DAML02
Galactic OC catalogue treats it as an OC with a diameter of 31
arcmin. The VVV NIR data exhibit an obvious OC main sequence (see
Fig. \ref{c2f2.fig}). \citet{Russeil12} obtained a distance $d = 1.72
\pm 0.26$ kpc and an extinction of $A_V= 4.52 \pm 0.68$ mag using 11
OB stars. Our isochrone fits yield a distance, colour excess and age
of $\mu_{0} = 11.16 \pm 0.25$ mag ($d = 1.7 \pm 0.2$ kpc),
$E(J-H)=0.28 \pm 0.02$ mag and $\log (t \mbox{ yr}^{-1}) = 7.7 \pm
0.3$, respectively. A distance modulus of $\mu_{0} = 11.17 \pm 0.38$
mag is determined from the $V$-band PLR if we adopt the V0470 Sco mean
magnitude of $m_{V}=11.0$ mag and $A_{V}= 4.52$ mag. The Cepheid V0470
Sco -- $\log P = 1.211$ [days] -- is located at roughly 70 arcmin from
the centre of NGC 6334, which is well beyond the catalogued size of
this cluster. However, a {\sl Herschel} image \citep{Russeil13} shows
that the cluster's structure extends to a radius of approximately 2
degrees. The radial velocity of V0470 Sco is $-3.4 \pm 0.5$ km
s$^{-1}$ \citep{Pejcha12}, which is comparable to both the H{\sc ii}
region's velocity \citep{Anderson11} and the mean velocity of the NGC
6334 complex \citep{Russeil13}. The proper motion of V0470 Sco,
$(\mu_{\alpha,{\rm Cep}},\mu_{\delta,{\rm Cep}}) = (-2.7 \pm 1.0,
1.1\pm 2.0)$ mas yr$^{-1}$, is also in the same range as that of NGC
6334, $(\mu_{\alpha,{\rm cl}},\mu_{\delta,{\rm cl}}) = (-1.2 \pm 3.1,
-0.2 \pm 3.4)$ mas yr$^{-1}$. Since Cepheids obey PLR and
  mass--luminosity relations, the long period of V0470 Sco implies and
  intrinsically high luminosity, high mass and young
  age. \citet{Bono05} provided a straightforward means to calculate
  Cepheid ages based on theoretical pulsation modelling; they obtained
  the period--age relation $\log( t \mbox{ yr}^{-1}) = (8.31 \pm 0.08)
  - (0.67 \pm 0.01) \log (P \mbox{ day}^{-1})$ for fundamental-mode
  Cepheids. Therefore, the age of V0470 Sco is $\log (t \mbox{
  yr}^{-1}) = 7.5 \pm 0.1$. The difference compared with the cluster
age is also within the $1\sigma$ age error (0.3 dex). Considering all
of these constraints, V0470 Sco is a high-confidence OC Cepheid.

Teutsch 14a is a small OC with a size of 7 arcmin \citep{Dias02}. The
Cepheid ASAS J180342$-$2211.0 is projected onto the cluster
centre. Based on the OC's CMD, ASAS J180342$-$2211.0 is located inside
the cluster's instability strip. Teutsch 14a is located in the
Galactic bulge and poorly studied. \citet{Schultheis14} estimated the
extinction values of and distances to eight regions inside the 7
arcmin OC radius, resulting in average values of $E(J-H)=0.51 \pm
0.10$ mag and $d = 2.0 \pm 0.5$ kpc. We determined the cluster's most
likely distance, colour excess and age using isochrone fitting, i.e.,
$\mu_{0} = 11.53 \pm 0.23$ mag ($d = 2.0 \pm 0.2$ kpc), $E(J-H)=0.48
\pm 0.01$ mag and $\log (t \mbox{ yr}^{-1}) = 7.5 \pm 0.3$. ASAS
J180342$-$2211.0 has a pulsation period of $\log P = 1.623$ [days]. It
is very important in the context of the PLR, since it is one of the
longest-period OC Cepheids known in the Galaxy. The Cepheid's age is
$\log (t \mbox{ yr}^{-1}) = 7.2 \pm 0.1$, which is within the range
inferred for the cluster's age. The proper motion of Teutsch 14a is
$(\mu_{\alpha,{\rm cl}},\mu_{\delta,{\rm cl}}) = (-0.2 \pm 3.9, 0.4
\pm 4.2)$ mas yr$^{-1}$, which is comparable to the values determined
by \citet{Dias14}. The Cepheid's proper motion is $(\mu_{\alpha,{\rm
    Cep}}, \mu_{\delta,{\rm Cep}}) = (-0.6 \pm 3.3, 5.9 \pm 2.5)$ mas
yr$^{-1}$, just inside the 1$\sigma$ boundary of cluster's
proper-motion distribution. Although radial-velocity data are not
available, ASAS J180342$-$2211.0 has all attributes to make it a
high-confidence OC Cepheid.

Majaess 170 is a small OC with a size of 5 arcmin \citep{Dias02}. ASAS
J160125$-$5150.3 is located 7 arcmin from the cluster centre. This
cluster has not been studied previously. The distance modulus, colour
excess and age we determined are $\mu_{0} = 11.30 \pm 0.28$ mag ($d =
1.8 \pm 0.3$ kpc), $E(J-H)=0.38 \pm 0.03$ mag and $\log (t \mbox{
  yr}^{-1}) = 8.1 \pm 0.2$, respectively. The proper motions of this
OC--Cepheid pair are $(\mu_{\alpha,{\rm cl}},\mu_{\delta,{\rm cl}} =
(-3.2 \pm 3.1, -3.1 \pm 3.3)$ mas yr$^{-1}$ and $(\mu_{\alpha,{\rm
    Cep}}, \mu_{\delta,{\rm Cep}}) = (-1.3 \pm 1.8, -0.2 \pm 1.8)$ mas
yr$^{-1}$. The Cepheid's proper-motion and distance-modulus
measurements imply that it is a high-probability cluster member. The
age of ASAS J160125$-$5150.3, derived from its period, is
approximately $\log (t \mbox{ yr}^{-1}) = 7.8 \pm 0.1$, which implies
a difference with respect to the host cluster of less than 0.3 dex.

Finally, Teutsch 77 is a small OC with a size of 7 arcmin
\citep{Dias02}. BB Cen is located 15 arcmin from the cluster
centre. The Cepheid's proper motion, $(\mu_{\alpha,{\rm Cep}},
\mu_{\delta,{\rm Cep}}) = (-3.9 \pm 1.3, 1.0 \pm 1.1)$ mas yr$^{-1}$,
is within the 1$\sigma$ range of the cluster's average proper motion,
$(\mu_{\alpha,{\rm cl}}, \mu_{\delta,{\rm cl}}) = (-4.2 \pm 4.2, -2.0
\pm 3.8)$ mas yr$^{-1}$. We obtained $\mu_{0} = 12.40 \pm 0.25$ mag
($d = 2.6 \pm 0.3$ kpc), $E(J-H)=0.10 \pm 0.02$ mag and $\log (t
\mbox{ yr}^{-1}) = 8.1 \pm 0.2$. A comparison of the cluster and
Cepheid properties suggests that BB Cen is an OC Cepheid.

\begin{table*}
\tiny
 \begin{minipage}{200mm}
\caption{Properties of our sample of 31 OCs and Cepheids.\label{c2table1}}
\begin{tabular}{@{}lcccclcccccc@{}}
  \hline
Cluster & $E(J-H)$ & $\log( P )$ & $\log( t )$ & \multicolumn{1}{c}{$\mu_{0}$} & Cepheid  & $\langle J \rangle$ & $\langle H \rangle$ & $\langle K_{\rm s} \rangle$ & $M_J$ & $M_H$ & $M_{K{\rm s}}$ \\
          & (mag)    & [day]        & [yr]  & \multicolumn{1}{c}{(mag)}     &          & (mag) & (mag) & (mag)  & (mag) & (mag) & (mag) \\
  \hline

Dolidze 34   & $0.36 \pm 0.02 $ & 1.043 & $7.9\pm0.2$ & $ 11.85 \pm 0.25 $& TY Sct         &7.23(0.01)&6.62(0.01)&6.40(0.01)& --5.57(0.27)& --5.82(0.25)& --5.83(0.23)\\
Dolidze 34   & $0.36 \pm 0.02 $ & 1.000 & $7.9\pm0.2$ & $ 11.85 \pm 0.25 $& CN Sct         &7.82(0.01)&7.06(0.01)&6.73(0.01)& --4.98(0.27)& --5.38(0.25)& --5.50(0.23)\\
Ruprecht 65  & $0.2  \pm 0.03 $ & 0.495 & $8.0\pm0.2$ & $ 11.07 \pm 0.38 $& AP Vel         &7.81(0.05)&7.36(0.09)&7.18(0.05)& --3.79(0.43)& --4.04(0.44)& --4.10(0.38)\\
Dolidze 52   & $0.24 \pm 0.02 $ & 0.808 & $7.6\pm0.3$ & $ 10.77 \pm 0.25 $& XX Sgr         &6.44(0.02)&5.98(0.02)&5.82(0.02)& --4.96(0.27)& --5.18(0.25)& --5.20(0.24)\\
NGC 6683     & $0.18 \pm 0.01 $ & 0.870 & $8.2\pm0.2$ & $ 11.62 \pm 0.23 $& CK Sct         &7.38(0.01)&6.81(0.01)&6.63(0.01)& --4.72(0.24)& --5.11(0.23)& --5.19(0.22)\\
Berkeley 58  & $0.23 \pm 0.02 $ & 0.640 & $8.0\pm0.2$ & $ 12.39 \pm 0.35 $& CG Cas         &8.95(0.09)&8.44(0.05)&8.26(0.03)& --4.05(0.44)& --4.33(0.38)& --4.38(0.36)\\
NGC 7790     & $0.14 \pm 0.02 $ & 0.688 & $7.9\pm0.2$ & $ 12.63 \pm 0.35 $& CF Cas         &8.60(0.01)&8.11(0.01)&7.94(0.01)& --4.41(0.37)& --4.75(0.35)& --4.84(0.33)\\
NGC 6087     & $0.06 \pm 0.005$ & 0.989 & $8.0\pm0.2$ & $ 9.84  \pm 0.11 $& S Nor          &4.68(0.02)&4.30(0.02)&4.15(0.02)& --5.32(0.13)& --5.64(0.13)& --5.76(0.13)\\
IC 4725      & $0.18 \pm 0.01 $ & 0.829 & $7.8\pm0.2$ & $ 8.82  \pm 0.13 $& U Sgr          &4.52(0.01)&4.09(0.01)&3.95(0.01)& --4.78(0.14)& --5.03(0.13)& --5.06(0.12)\\
vdBergh 1    & $0.26 \pm 0.015$ & 0.731 & $7.9\pm0.2$ & $ 11.11 \pm 0.24 $& CV Mon         &7.31(0.01)&6.77(0.01)&6.57(0.01)& --4.49(0.25)& --4.77(0.24)& --4.82(0.23)\\
NGC 129      & $0.165\pm 0.015$ & 0.903 & $7.8\pm0.2$ & $ 11.16 \pm 0.24 $& DL Cas         &6.56(0.01)&6.09(0.01)&5.93(0.01)& --5.04(0.25)& --5.35(0.24)& --5.41(0.23)\\
Collinder 394& $0.095\pm 0.005$ & 0.822 & $8.0\pm0.2$ & $ 9.35  \pm 0.16 $& BB Sgr         &5.05(0.02)&4.66(0.02)&4.50(0.02)& --4.55(0.18)& --4.84(0.18)& --4.95(0.18)\\
Turner 2     & $0.18 \pm 0.01 $ & 1.339 & $8.0\pm0.2$ & $ 11.22 \pm 0.13 $& WZ Sgr         &5.26(0.01)&4.74(0.01)&4.57(0.01)& --6.44(0.14)& --6.78(0.13)& --6.84(0.12)\\
Teutsch 14a  & $0.48 \pm 0.01 $ & 1.623 & $7.5\pm0.3$ & $ 11.53 \pm 0.23 $& 180342$-$2211.0&5.59(0.07)&4.76(0.07)&4.28(0.06)& --7.21(0.29)& --7.56(0.29)& --7.76(0.27)\\
Majaess 170  & $0.38 \pm 0.03 $ & 0.699 & $8.1\pm0.2$ & $ 11.30 \pm 0.28 $& 160125$-$5150.3&7.97(0.02)&7.22(0.04)&6.90(0.03)& --4.33(0.30)& --4.70(0.29)& --4.79(0.26)\\
Lynga 6      & $0.38 \pm 0.02 $ & 1.033 & $7.9\pm0.2$ & $ 11.66 \pm 0.25 $& TW Nor         &7.45(0.02)&6.72(0.02)&6.39(0.02)& --5.25(0.27)& --5.58(0.25)& --5.68(0.24)\\
Teutsch 77   & $0.1  \pm 0.02 $ & 0.757 & $8.1\pm0.2$ & $ 12.04 \pm 0.25 $& BB Cen         &8.00(0.02)&7.57(0.02)&7.39(0.02)& --4.30(0.27)& --4.63(0.25)& --4.76(0.24)\\
NGC 6334     & $0.28 \pm 0.02 $ & 1.211 & $7.7\pm0.3$ & $ 11.16 \pm 0.25 $& V0470 Sco      &5.95(0.07)&5.21(0.07)&4.87(0.06)& --5.95(0.32)& --6.41(0.30)& --6.59(0.28)\\
Kharchenko 3 & $0.26 \pm 0.025$ & 0.744 & $8.1\pm0.2$ & $ 11.51 \pm 0.37 $& 182714$-$1507.1&7.89(0.05)&7.32(0.04)&7.14(0.04)& --4.31(0.42)& --4.62(0.38)& --4.65(0.36)\\
ASCC 61      & $0.08 \pm 0.01 $ & 0.687 & $8.0\pm0.2$ & $ 11.19 \pm 0.23 $& SX Car         &7.26(0.02)&6.88(0.02)&6.74(0.02)& --4.14(0.25)& --4.44(0.24)& --4.53(0.23)\\
ASCC 61      & $0.08 \pm 0.01 $ & 1.277 & $8.0\pm0.2$ & $ 11.19 \pm 0.23 $& VY Car         &5.41(0.02)&4.97(0.02)&4.79(0.02)& --5.99(0.25)& --6.35(0.24)& --6.48(0.23)\\
ASCC 69      & $0.06 \pm 0.01 $ & 0.985 & $8.0\pm0.2$ & $ 9.74  \pm 0.23 $& S Mus          &4.50(0.02)&4.15(0.02)&4.00(0.02)& --5.40(0.25)& --5.69(0.24)& --5.80(0.23)\\
Collinder 220& $0.12 \pm 0.015$ & 0.728 & $8.1\pm0.2$ & $ 11.48 \pm 0.24 $& UW Car         &7.31(0.02)&6.92(0.02)&6.67(0.02)& --4.49(0.26)& --4.76(0.24)& --4.94(0.24)\\
Feinstein 1  & $0.12 \pm 0.01 $ & 1.589 & $7.0\pm0.3$ & $ 10.88 \pm 0.23 $& U Car          &4.14(0.02)&3.69(0.02)&3.51(0.02)& --7.06(0.25)& --7.39(0.24)& --7.50(0.23)\\
Trumpler 35  & $0.33 \pm 0.01 $ & 1.294 & $7.8\pm0.2$ & $ 11.43 \pm 0.18 $& RU Sct         &5.89(0.01)&5.27(0.01)&5.07(0.01)& --6.41(0.19)& --6.70(0.18)& --6.71(0.17)\\
Ruprecht 175 & $0.065\pm 0.01 $ & 1.215 & $7.8\pm0.2$ & $ 10.23 \pm 0.18 $& X Cyg          &4.42(0.02)&3.98(0.02)&3.82(0.02)& --5.98(0.20)& --6.36(0.19)& --6.48(0.18)\\
NGC 5662     & $0.1  \pm 0.01 $ & 0.740 & $8.0\pm0.2$ & $ 9.24  \pm 0.13 $& V Cen          &5.02(0.02)&4.65(0.02)&4.49(0.02)& --4.48(0.15)& --4.75(0.14)& --4.85(0.13)\\
Alessi 95    & $0.08 \pm 0.02 $ & 0.440 & $8.0\pm0.2$ & $ 8.04  \pm 0.20 $& SU Cas         &4.49(0.02)&4.22(0.02)&4.11(0.02)& --3.77(0.22)& --3.95(0.20)& --4.01(0.19)\\
NGC 1647     & $0.1  \pm 0.01 $ & 0.651 & $8.2\pm0.2$ & $ 8.74  \pm 0.13 $& SZ Tau         &4.78(0.02)&4.43(0.02)&4.30(0.02)& --4.22(0.15)& --4.47(0.14)& --4.54(0.13)\\
NGC 6649     & $0.42 \pm 0.01 $ & 0.651 & $8.1\pm0.2$ & $ 10.89 \pm 0.13 $& V0367 Sct      &7.66(0.03)&6.98(0.04)&6.67(0.04)& --4.34(0.16)& --4.60(0.16)& --4.66(0.15)\\
NGC 6067     & $0.12 \pm 0.02 $ & 1.053 & $8.0\pm0.2$ & $ 11.22 \pm 0.25 $& V0340 Nor      &6.26(0.01)&5.73(0.01)&5.60(0.01)& --5.29(0.27)& --5.68(0.25)& --5.79(0.23)\\

\hline
 \end{tabular}
\end{minipage}
\end{table*}

\section{Near-Infrared Period--Luminosity Relations}

In attempting to establish Cepheid NIR PLRs, we run into the problem
that only some Cepheids have complete NIR light
curves. \citet{Monson11} published $JHK$ photometry for 131 northern
Galactic classical Cepheids, while $JHK$ mean magnitudes in the South
African Astronomical Observatory (SAAO) system are available for 223
southern Cepheids \citep{van Leeuwen07}. The mean NIR magnitudes have
been measured directly for only 19 OC Cepheids \citep{Chen15}. To
supplement these data sets with OC Cepheids without readily available
NIR light curves, we need to transform single-epoch NIR magnitudes to
mean magnitudes. Fortunately, Cepheid NIR light curves are related to
their $V$-band light curves, so that we can estimate the mean NIR
magnitudes based on the corresponding $V$-band light curves
\citep{Soszynski05}. The amplitudes of Cepheid light curves decrease
from the optical to the NIR regime. The NIR-to-optical amplitude ratio
is around 0.3--0.4. These considerations allow us to derive the OC
Cepheids' NIR amplitudes. We calculated the phases of the
  single-epoch NIR magnitudes and combined them with the seven Fourier
  coefficients of the light-curve template from
  \citet{Soszynski05}. This light-curve template was developed based
  on 30 Cepheids in the Galaxy and 31 Cepheids in the Large Magellanic
  Cloud (LMC) with complete $VIJHK_{\rm s}$ light curves. The
  $JHK_{\rm s}$ light-curve template was developed with respect to the
  $V$-band light curve. The resulting differences between the derived
and real mean magnitudes are less than 0.03 mag, which is acceptable
because this level of uncertainty is small compared with the
uncertainty in the distance modulus. We thus obtained $JHK_{\rm s}$
mean magnitudes from 2MASS single-epoch photometry for seven
additional Cepheids without directly measured NIR mean magnitudes,
including AP Vel, CG Cas, V0367 Sct, V0470 Sco, ASASJ182714$-$1507.1,
ASAS J180342$-$2211.0 and ASAS J160125$-$5150.3 (see Table
\ref{c2table3}).

%\begin{figure}
%\centering
%\includegraphics[width=80mm]{c2f4.eps}
%\caption{Comparison of converted NIR mean magnitudes ($M$) with real
%  NIR mean magnitudes ($\langle M \rangle$). We have used 19 of the 31
%  Cepheids for which both magnitudes are available.} \label{c2f4.fig}
%\end{figure}

\begin{table*}
\caption{Mean magnitudes of seven Cepheids. $J(\varphi)$,
  $H(\varphi)$ and $K_{\rm s}(\varphi)$ are 2MASS magnitudes at phase
  $\varphi$, while $\langle J \rangle$, $\langle H \rangle$ and
  $\langle K_{\rm s} \rangle$ are the derived mean magnitudes.
\label{c2table3}}
\begin{tabular}{@{}lcccccccc@{}}
  \hline
 Cepheid & Amp$(V)$ & Phase ($\varphi$) & $J(\varphi)$ & $H(\varphi)$ &  $K_{\rm s}(\varphi)$ & $\langle J \rangle$ & $\langle H \rangle$ & $\langle K_{\rm s} \rangle$ \\
          & (mag)    &        &  (mag) & (mag) & (mag)  & (mag) & (mag) & (mag) \\
  \hline

V0367 Sct       & 0.54 & 0.42 &7.716(0.024)&7.004(0.027) & 6.681(0.021) &7.657(0.034)&6.983(0.043)&6.671(0.038)\\
CG CAS          & 0.27 & 0.84 &8.832(0.044)&8.313(0.033) & 8.136(0.021) &8.954(0.086)&8.435(0.050)&8.257(0.034)\\
AP Vel          & 0.00 & 0.72 &7.730(0.026)&7.351(0.059) & 7.188(0.017) &7.807(0.052)&7.355(0.090)&7.181(0.048)\\
180342$-$2211.0 & 0.84 & 0.76 &5.658(0.021)&4.880(0.042) & 4.403(0.033) &5.590(0.065)&4.764(0.069)&4.277(0.057)\\
160125$-$5150.3 & 0.06 & 0.30 &7.921(0.018)&7.205(0.026) & 6.889(0.020) &7.968(0.023)&7.223(0.036)&6.902(0.031)\\
V0470 Sco       & 0.95 & 0.92 &5.900(0.027)&5.245(0.024) & 4.919(0.023) &5.949(0.072)&5.209(0.066)&4.869(0.064)\\
182714$-$1507.1 & 0.79 & 0.64 &7.977(0.020)&7.415(0.020) & 7.227(0.024) &7.894(0.054)&7.324(0.036)&7.136(0.037)\\

\hline
 \end{tabular}
 \end{table*}

\begin{figure*}
\begin{minipage}{180mm}
\centering
\includegraphics[width=180mm]{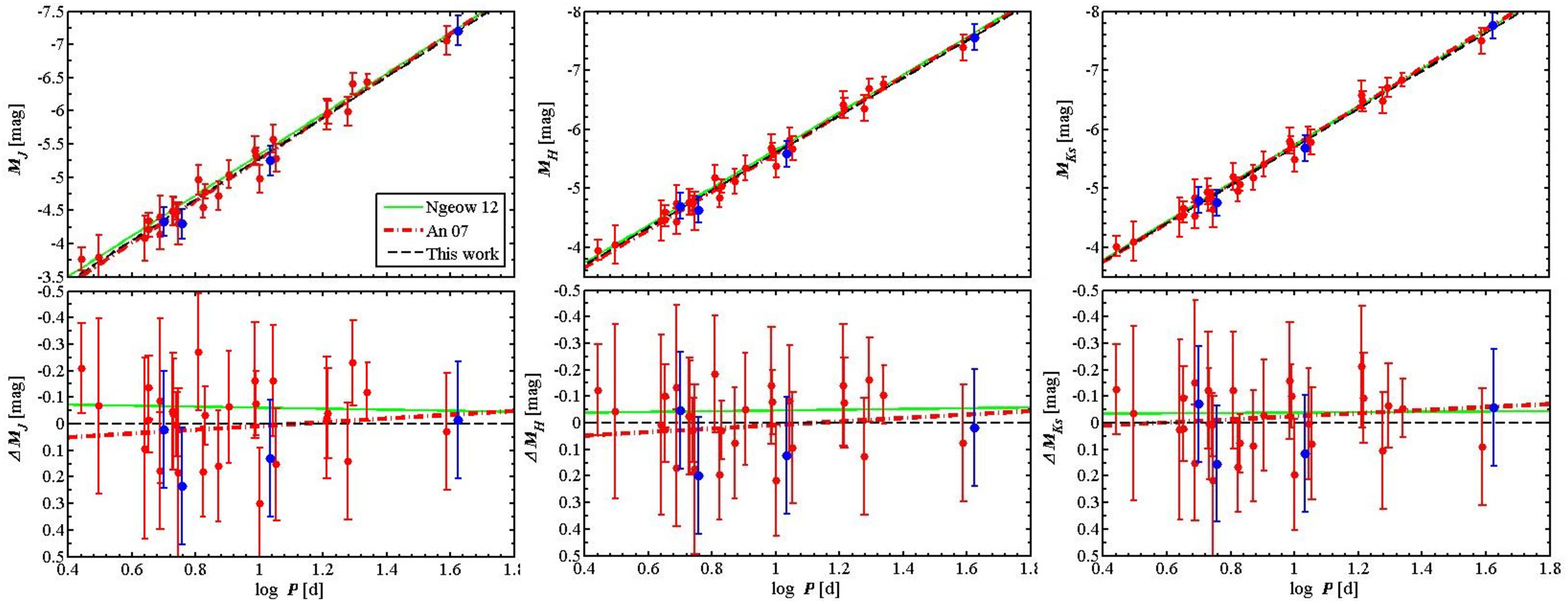}
\caption{(Top) $JHK_{\rm s}$ PLRs for 31 OC Cepheids. Red bullets:
  confirmed OC Cepheids; blue bullets: new, high-confidence OC
  Cepheids found in this paper. The green solid line is the $JHK_{\rm
    s}$ PLR for Galactic Cepheids \citep{Ngeow12}; the red dotted line
  is from \citet{An07} and the black dashed line is our linear
  fit. (Bottom) Differences between the three PLRs, where our newly
  determined PLR is used as reference.\label{c2f3.fig}}
\end{minipage}
\end{figure*}

Establishing and calibrating PLRs is vital for the use of Cepheids as
distance tracers, especially when using multiple passbands in the
NIR. The use of OCs to derive PLRs is a useful and independent
method. \citet{Tammann03} derived $BVI$ PLRs for 25 OC Cepheids,
\citet{An07} obtained $BVI_{\rm c}JHK_{\rm s}$ PLRs based on seven
well-studied OC Cepheids and \citet{Chen15} obtain a $J$-band PLR
based on 19 OC Cepheids. In this paper, we determined the $JHK_{\rm
  s}$ PLRs based on 31 OC Cepheids. This represents a significant
improvement in the $J$-band PLR of \citet{Chen15}, while statistically
meaningful $H$- and $K_{\rm s}$-band OC--Cepheid PLRs were determined
for the first time:
\begin{equation}\label{c2equation1}
  \begin{aligned}
   &\langle M_J \rangle = (-3.077 \pm 0.090) [\log P-1]-(5.279 \pm 0.028), \sigma_J=0.148; \\
   &\langle M_H \rangle = (-3.167 \pm 0.075) [\log P-1]-(5.601 \pm 0.023), \sigma_H=0.124; \\
   &\langle M_{K{\rm s}} \rangle = (-3.224 \pm 0.073) [\log P-1]-(5.693 \pm 0.022), \sigma_{K{\rm s}}=0.120.
   \end{aligned}
\end{equation}
These are currently the most accurate PLRs based on the largest OC
Cepheid sample available to date, characterized by an accuracy that is
improved by 40 per cent compared with previous determinations. The
PLRs derived here agree well with results based on other methods
\citep{An07,Ngeow12}, considering the uncertainties: see
Fig. \ref{c2f3.fig}. The $J$-band PLRs exhibit obvious zero-point
differences. These zero-point differences become smaller from $J$ to
$K_{\rm s}$, and in the $K_{\rm s}$ band the differences among the
three relations are always less than 0.06 mag. In addition, for the
$J$-band PLRs, we find at the short-period end -- $\log P \simeq 0.4$
[days] -- a 0.14 mag difference between \citet{An07} and
\citet{Ngeow12}, while our PLR yields a compromise value. At the
long-period end -- $\log P \simeq 1.6$ [days] -- these differences
disappear. This is owing to the small number of long-period
(high-mass) Cepheids, which play an important role in constraining the
slope of the PLR. The slopes we have determined here are
  comparable with those of \citet{Ngeow12}. With respect to the slopes
  determined by \citet{An07}, however, we find differences of 0.08,
  0.06 and 0.06 mag in the $J$, $H$ and $K_{\rm s}$ bands,
  respectively. Since our PLR slopes and those of \citet{Ngeow12} were
  derived based on Galactic Cepheids, while the slopes of \citet{An07}
  used the PLR slopes of Cepheids in the LMC as their basis, this
  suggests that the PLR slopes may be different for the LMC and the
  Galaxy, even at NIR wavelengths. However, these differences may also
  have been caused by the application of different methods by their
  respective authors. The slopes and intercepts of the three sets of
  $JHK_{\rm s}$ PLRs are listed in Table \ref{c2table4}. In view of
  the prevailing uncertainties, we cannot firmly conclude that the
  differences cited above are statistically significant. Additional
  Cepheid samples are needed to constrain the statistical
  uncertainties.

\begin{table*}
\caption{Slopes ($a$) and intercepts ($b$) of the NIR PLRs derived by
  \citet{An07}, \citet{Ngeow12} and in this paper. The PLRs are
  defined as $\langle M \rangle = a[\log P-1]+b$, where $M$ are the
  intrinsic magnitudes and $P$ the Cepheids' periods.\label{c2table4}}
\begin{tabular}{@{}lcccccccc@{}}
  \hline
              &         $a(J)$    &      $b(J)$       &      $a(H)$       &     $b(H)$        & $a(K_{\rm s})$  & $b(K_{\rm s})$\\
  \hline

An07           & $-3.148\pm0.053$ & $-5.271\pm0.076$ & $-3.233\pm0.044$ & $-5.593\pm0.069$ & $-3.282\pm0.040$ & $-5.718\pm0.064$\\
Ngeow12        & $-3.058\pm0.021$ & $-5.340\pm0.019$ & $-3.181\pm0.022$ & $-5.648\pm0.020$ & $-3.231\pm0.021$ & $-5.732\pm0.020$\\
This paper     & $-3.077\pm0.090$ & $-5.279\pm0.028$ & $-3.167\pm0.075$ & $-5.601\pm0.023$ & $-3.224\pm0.073$ & $-5.693\pm0.022$\\

\hline
 \end{tabular}
 \end{table*}

\section{Conclusion}

The availability of VVV DR1 provides a unique opportunity to uncover
and study faint and highly obscured OCs in the Galactic plane near the
Galactic Centre. Using the DAML02 OC catalogue and other new OCs found
in the VVV data, we have carefully examined 22 OCs and discovered four
new OC Cepheids. Parameters such as distances, reddening values and
ages were better determined compared with previous work based on
isochrone fitting. By comparison of distances, apparent positions,
proper motions and ages of the clusters and Cepheids, the newly found
OC--Cepheid pairs include NGC 6334 and V0470 Sco, Teutsch 14a and ASAS
J180342$-$2211.0, Majaess 170 and ASAS J160125$-$5150.3, and Teutsch
77 and BB Cen. ASAS J180342$-$2211.0 is the longest-period Cepheid --
$\log P = 1.623$ [days] -- thus far found in an OC, which is important
in the context of constraining the slope of the PLR.

The currently most complete NIR Cepheid PLRs based on 31 OC Cepheids
were obtained (a significant improvement from the previous PLRs which
were based on up to 19 OC Cepheids), i.e., $\langle M_J \rangle =
(-3.077 \pm 0.090) \log P-(2.202 \pm 0.090)$, $\langle M_H \rangle =
(-3.167 \pm 0.075) \log P-(2.434 \pm 0.074)$ and $\langle M_{K{\rm s}}
\rangle = (-3.224 \pm 0.073) \log P-(2.469 \pm 0.072)$. The associated
uncertainties have been improved by 40 per cent compared with previous
Cepheid PLRs based on the OC-matching method. Our PLRs are in good
agreement with the best NIR PLRs available for all Galactic Cepheids
obtained using other methods. Particularly in the $K_{\rm s}$ band,
the zero-point differences are very small; differences of less than
0.06 mag are expected for any Cepheid.

Since there are more than 20,000 Cepheids and 20,000 OCs in the
Galaxy, we expect that thousands of OC Cepheids will be discovered in
the future. {\sl Gaia} will observe some 9000 Cepheids
\citep{Windmark11}, which will improve the distances, proper motions
and radial velocities of not only the Cepheids but also their host
clusters. By combining the more accurate parallax distances from {\sl
  Gaia} with OC distances based on isochrone fitting, the Cepheid PLRs
will be left with only small systematic errors.

We thank the referees for their comments. We are grateful
for support from the National Natural Science Foundation of China
through grants 11633005, U1631102, 11373010 and 11473037.

\end{document}